\documentclass[twocolumn]{aastex631}
\usepackage[fleqn]{amsmath}
\usepackage{mathrsfs}
\usepackage[french]{babel}

\shorttitle{Early stages of Galilean moon formation in a water-depleted environment}
\shortauthors{Mousis et al.}
\graphicspath{{./}{figures/}}

\begin{document}

\title{Early stages of Galilean moon formation in a water-depleted environment}

\author[0000-0001-5323-6453]{Olivier Mousis}
\affiliation{Aix-Marseille Universit\'e, CNRS, CNES, Institut Origines, LAM, Marseille, France}
\affiliation{Institut universitaire de France (IUF)}
\author[0000-0002-3289-2432]{Antoine Schneeberger}
\affiliation{Aix-Marseille Universit\'e, CNRS, CNES, Institut Origines, LAM, Marseille, France}
\author[0000-0003-2279-4131]{Jonathan I. Lunine}
\affiliation{Department of Astronomy, Cornell University, Ithaca, NY, USA}
\author[0000-0002-2161-4672]{Christopher R. Glein}
\affiliation{Space Science Division, Southwest Research Institute, San Antonio, TX, USA}
\author[0000-0001-8262-9678]{Alexis Bouquet}
\affiliation{Aix-Marseille Université, CNRS, Institut Origines, PIIM, Marseille, France}
\affiliation{Aix-Marseille Universit\'e, CNRS, CNES, Institut Origines, LAM, Marseille, France}
\author[0000-0002-4242-3293]{Steven D. Vance}
\affiliation{Jet Propulsion Laboratory, California Institute of Technology, 4800 Oak Grove Dr, Pasadena, CA 91109-8001, USA}



\begin{abstract}

A key feature of the Galilean satellite system is its monotonic decrease in bulk density with distance from Jupiter, indicating an ice mass fraction that is zero in the innermost moon Io, and about half in the outer moons Ganymede and Callisto. Jupiter formation models, and perhaps the Juno spacecraft water measurements, are consistent with the possibility that the Jovian system may have formed, at least partly, from ice-poor material. And yet, models of the formation of the Galilean satellites usually assume abundant water ice in the system. Here, we investigate the possibility that the Jovian circumplanetary disk was populated with ice-depleted chondritic minerals, including phyllosilicates. We show that the dehydration of such particles and the outward diffusion of the released water vapor allow condensation of significant amounts of ice in the formation region of Ganymede and Callisto in the Jovian circumplanetary disk. Our model predicts that Europa, Ganymede and Callisto should have accreted little if any volatiles other than water ice, in contrast to the comet-like composition of Saturn's moon Enceladus. This mechanism allows for the presence of ice-rich moons in water-depleted formation environments around exoplanets as well.

\end{abstract}



\section{Introduction} 
\label{sec:intro}

The amount of water that was available to form Jupiter and its regular satellite system is still under debate. Recent formation scenarios of Jupiter invoke a substantial migration of the planet during its growth, perhaps interior to the location of the snowline in the protosolar nebula (PSN) \citep{Ob19,Sc21,Sh22}. Those scenarios are in agreement with the 2-$\sigma$ error bar associated with the Juno measurement of the deep water abundance performed at a single latitude near the equator of Jupiter, leaving open the possibility it could be subsolar \citep{Li20}, and that a global depletion cannot be excluded \citep{He22}. Hence, it is plausible that, depending on Jupiter's position on its migration path in the PSN, the solids accreted by the growing planet and/or by its late-forming circumplanetary disk (CPD) \citep{Sz17} were no richer in water than carbonaceous chondrites. This idea of forming the moons from ice-poor material is supported by recent interior evolution models of Europa suggesting that it could have accreted from carbonaceous chondritic minerals, including phyllosilicates such as serpentine \citep{Me21}. The high temperatures required in Europa to form an iron-rich core would have been sufficient to release volatiles from the rocks and constitute its present-day hydrosphere, which does not exceed $\sim$8 wt\% \citep{Me21}. If Europa accreted from ice-free building blocks that were not partly devolatilized in the Jovian CPD, a fundamental question is to understand how both Ganymede and Callisto can be made of a roughly equal mix of ice and refractory material \citep{So02}, assuming the subdisk was only fueled from similar material.

Here, we propose a scenario based on the properties of diffusive redistribution of water vapor throughout the CPD to explain how the Galilean moons could accrete building blocks with the observationally constrained ice-to-rock ratio during their formation, even though the {CPD was being fed} with ice-free minerals. In such a picture, Io would have accreted from the {dehydrated product} of phyllosilicate--rich solids, subsequent to their inward drift beyond the formation location of Europa in the CPD. {Phyllosilicates form a large group of hydrous minerals, most notably serpentines and smectites. They contain OH as part of their crystal structures and can incorporate H$_2$O molecules between layers in their structures. Carbonaceous chondrites are often rich in phyllosilicates, which are responsible for the high content of mineral-bound water of $\sim$10 wt$\%$ in CI/CM chondrites \citep{Al19}.} Phyllosilicate--rich solids drifting within the CPD could either result from the disruption of a first generation of parent bodies that underwent aqueous alteration, or from the capture of micron-sized grains that had been previously hydrated in the PSN \citep{Ci05}. This formation mechanism invoked for Io is similar to those previously mentioned in the literature in which the moon would have formed from devolatilized solids \citep{Ca02,Ro17}, but with the notable exception that the loss of water happens inside the Phyllosilicate Dehydration Line (PDL) (at $\sim$400--600K, see below), instead of the snowline at $\sim$160 K \citep{Ha15,Sa10}. Forming Io from aqueously altered materials would allow some oxidation driven by H$_2$ escape, which would be consistent with inferences of the oxidation state of its interior \citep{Zo99}. A key consequence of water vapor release at the PDL is its diffusive redistribution throughout the CPD. Water vapor diffusing outward beyond the PDL is susceptible to condensing again once it reaches the colder regions of the CPD, increasing the ice--to--rock ratio of the solids located beyond the snowline {(see Fig. \ref{fig1} for a representation of the relative positions of snowline and PDL in the CPD)}.  {The proposed scenario is consistent with either a subsolar or supersolar water abundance in Jupiter.}

\section{Disk and transport models} 
\label{sec:model}

To assess the extent of the water vapor distribution within the Jovian CPD and examine whether it can explain by itself the substantial ice content of Ganymede and Callisto, we used a standard one-dimensional gas-starved accretion disk model derived from the literature \citep{Ca02,Sa10,An21}. 

The disk is fed through its upper layers from its inner edge up to the centrifugal radius $R_c$ by gas and gas-coupled solids inflowing from the PSN. Our model considers the transport of porous grains and pebbles, from a few microns to the centimeter-size scale level, in addition to gas. The outer radius of the CPD is defined by $R_d$~=~150~Jovian radii ($R_{Jup}$), based on 3D hydrodynamic simulations \citep{Ta12}. The surface density of the CPD is given by \citep{Ca02}:

\begin{equation}
\Sigma_g \simeq \frac{4 F_p}{15 \pi \nu} \lambda (r),
\label{eq1}
\end{equation}

\noindent where $F_p$ is the total infall rate and $\nu$ is the turbulent viscosity of the disk gas. The turbulent viscosity of the CPD is defined by $\nu = \alpha C_s^2 / \Omega_K$ \citep{Sh73}, where the {viscosity parameter $\alpha$ is set to the canonical value of 10$^{-3}$ \citep{Ca02,Sa10}}. $\Omega_K$ = $\sqrt{GM_{Jup}/r^3}$ is the Keplerian frequency, where $G$ is defined as the gravitational constant, $M_{Jup}$ is the mass of Jupiter, and $r$ is the radial distance to Jupiter within the CPD. $C_s$ is the isothermal sound speed given by $Cs = \sqrt{kT/\mu m_p}$, where $k$ is the Boltzmann constant, $\mu$ the mean molecular weight, and $m_p$ the proton mass. The coefficient $\lambda(r)$, defined by \cite{Ca02} and \cite{Sa10}, is:

\begin{equation}
\begin{split}
\lambda(r) = \frac{5}{4}-\sqrt{\frac{R_c}{R_d}}-\frac{1}{4}\Big(\frac{r}{R_c}\Big)^2  ~{\rm for}~r<R_c \\
{\rm and~} \lambda(r) = \sqrt{\frac{R_c}{r}}- \sqrt{\frac{R_c}{R_d}}    ~{\rm for}~r \ge R_c,
\end{split}
\label{eq2}
\end{equation}

\noindent with the value of $R_c$ set to 30 $R_{Jup}$ \citep{Ca02,Sa10}. The total infall rate follows an exponential relationship via $F_p~=~F_{p,0}\exp(-t/\tau_{\text{disk}})$, where the timescale $\tau_{disk}$ taken by the PSN to deplete is set equal to 3$\times$ 10$^6$ yr \citep{Sa10}. $F_{p,0}$, which corresponds to the total infall rate during the steady accretion state of the PSN, is set to $5 \times 10^{-7}$ $M_{Jup}$~/~yr \citep{Sa10}. This value allows us to start our calculations at a relatively late stage of the CPD evolution since we are only concerned with the architecture of a satellite system formed by the last survivors \citep{Sa10}.

We assume that viscous dissipation is the main heat source in our CPD model, implying that the disk's midplane temperature profile $T_{d,0}$ at $t$~=~0 is derived as follows \citep{Sa10}:

\begin{equation}
T_{d,0}^4(r) = \frac{3 \Omega_K^2}{10 \pi \sigma_{sb}} F_{p,0}~\lambda(r),
\label{eq3}
\end{equation}

\noindent where $\sigma_{sb}$ is the Stefan-Boltzmann constant. With time, the CPD temperature decreases as:

\begin{equation}
T_{d} \simeq T_{d,0}\exp{\Big(\frac{-t}{4\tau_{disk}}\Big)}.
\label{eq4}
\end{equation}

In our model, particles are injected into the CPD with a uniform size $a~=~10^{-6}$ m and density $\rho_s~=~1~\text{g cm}^{-3}$. These porous and micron-sized phyllosillicate-rich grains evolve in size and position through collisions, fragmentation, and radial drift \citep{Bi12,An21}. In this approach, the size of grains increases before reaching an equilibrium corresponding to the minimum value between fragmentation and radial drift. Fragmentation occurs when the relative velocity of the dust grains due to turbulent motion exceeds the fragmentation velocity threshold $u_f$, which is set to 10 m s$^{-1}$ in our calculations \citep{Bi12,An21}. The size of the grains, limited by their fragmentation, is then:

\begin{equation}
a_{\text{frag}} = 0.37\frac{2\Sigma_g}{3\pi\rho_s\alpha}\frac{u_f^2}{c_s^2}.
\end{equation}

In many cases, the inward drift timescale of the grains is much smaller than that needed to {let them grow significantly.} Their size, limited by their drift, is given by:

\begin{equation}
a_{\text{drift}} = 0.55\frac{2\Sigma_d}{3\pi\rho_s}\frac{v_K^2}{c_s^2}\Big|\frac{\text{d}\ln P}{\text{d}\ln r}\Big|^{-1},
\end{equation}

\noindent with $v_K$ the Keplerian velocity, $P = c^2_s\rho_g$ and $\rho_g = \Sigma_g/2\pi H_g$ the pressure and gas density at midplane. The prefactor $f_d = 0.55$ represents the offset of the representative size with respect to the maximum attainable size of the dust grains \citep{Bi12}.  

We assume that, once the phyllosillicate-rich grains have crossed the PDL, water vapor is immediately released into the CPD. Water vapor then radially diffuses and advects, until part of it condenses when going beyond the snowline. $\Sigma_i$ represents the surface density of species $i$ investigated here, which is either in vapor (H$_2$O) or solid (phyllosilicate and/or crystallized H$_2$O) form. The advection-diffusion equation is integrated following a forward Euler integration \citep{Bi12}:

\begin{equation}
\frac{\Sigma_i}{\partial t} + \frac{1}{r}\frac{\partial}{\partial r} \Bigg[ r\Bigg(\Sigma_i v_i - D_i\Sigma_g\frac{\partial}{\partial r}\Bigg(\frac{\Sigma_i}{\Sigma_g}\Bigg)\Bigg)\Bigg] - \dot{Q} = 0,
\end{equation}
     
\noindent where $v_i$ and $D_i$ are the radial velocities and the diffusivities of species $i$ respectively. $\dot{Q}$ corresponds to the source term of H$_2$O vapor released to the gas, and is given by $\Sigma_\text{phyllosilicate}~\times~f_{\rm H_2O}~/~\Delta t$ beyond the PDL, with $f_{\rm H_2O}$ the fraction of H$_2$O in phyllosilicates and $\Delta t$ the timestep of the simulation. The last values to be calculated are the dust velocity $v_d$, the Stokes number $St$, the gas radial velocity, and the diffusivity \citep{Bi12}. $v_d$ is given by:

\begin{equation}
v_d = -\frac{2 \text{St}}{1+\text{St}}\eta v_K + \frac{1}{1+\text{St}^2}v_g,
\end{equation}

\noindent where $v_K$ is the Keplerian velocity, $v_g$ is the inward radial velocity of the gas given by $v_g~=~-~3\nu/2r$, with $\nu$ and $\eta$ the diffusivity and viscosity of the gas, respectively. The Stokes Number, which describes the aerodynamic properties of the particles, is determined as follows:

\begin{equation}
\text{St} = \frac{a\pi\rho_s}{2\Sigma_g}.
\end{equation}

\noindent The diffusivity $D$ of the vapor species is assumed to be that of the gas $D_g = \nu$ and the diffusivity of the dust is given by:

\begin{equation}
D_d = \frac{D_g}{1+\text{St}}.
\end{equation}

Finally, the prescription that injects gas and gas-coupled dust from the PSN to the CPD region inside the centrifugal radius $R_c$ is given by \cite{Ca02} and \cite{An21}: 

\begin{equation}
\dot{\Sigma}_\text{solids} =
\begin{cases}
\frac{f\times F_p}{\pi R_c^2}  &  [r<R_c] \\
0    & [r>R_c]
\end{cases},
\end{equation}

\noindent where $f$ is the solids-to-gas ratio in our CPD model, here assumed to be $\sim$3~$\times$~10$^{-3}$ in mass fraction, based on the abundance of silicates and other rocky materials estimated for the PSN \citep{Lo03}. By doing so, the gas component of the CPD sustains a quasi-steady state, and viscously spreads inward and outward from $R_c$. Solids also accumulate in the formation region of the moons, close to which they are initially delivered. Note that our model provides rather similar results when $u_f$ is assumed to 1 m s$^{-1}$ instead of 10 m s$^{-1}$, as postulated in this work.

Figure 1 represents the temperature, pressure, and surface density profiles of our CPD model at three different epochs of its evolution ($t$ = 10$^4$, 10$^5$, and 10$^6$ yr). The time evolution of these physical quantities is slow in the Jovian CPD, as illustrated by the limited inward motion of the snowline (set to 160 K \citep{Ha15,Sa10}), which remains in the {$\sim$23.6--25.9 $R_{Jup}$ region over 10$^6$ yr. This is due to the injection of gas at a very slowly decreasing rate within the position of $R_c$, implying that the disk can be considered as stationary over short timescales. 

\begin{figure}
\resizebox{\hsize}{!}{\includegraphics[angle=0]{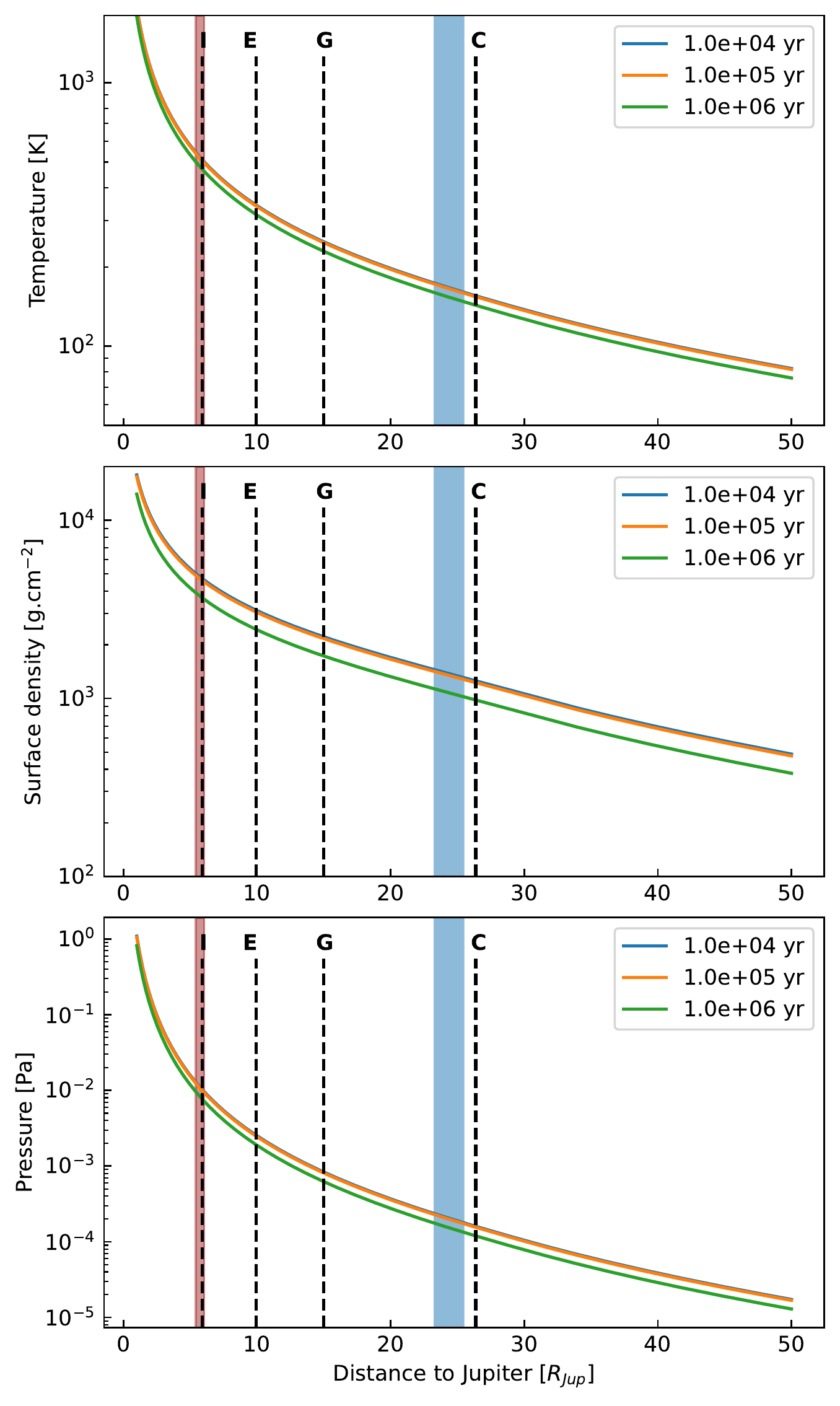}}
\caption{Temperature, surface density, and pressure profiles in the CPD expressed in units of Jovian radii ($R_{Jup}$), and calculated at $t$ = 10$^4$, 10$^5$, and 10$^6$ yr. The vertical bars designated by the letters I, E, G and C correspond to the current orbits of Io, Europa, Ganymede and Callisto, {respectively. The} vertical brown and blue rectangles correspond to the time evolution of the PDL and snowline (SL) over 10$^6$ yr in the CPD, respectively. A phyllosilicate dehydration temperature of 500 K is assumed in the figure.}
\label{fig1}
\end{figure}

\section{Results} 
\label{sec:results}

Figure 2 displays the evolution over one million years of the water vapor and ice abundance profiles normalized to the initial abundance in the phyllosilicate--bearing solids (10 wt\%; \cite{Al19}). Phyllosilicate--rich grains and pebbles drift inward within the CPD, and release the trapped water as vapor once they cross the PDL. It is assumed that the diffusion timescale of the released water vapor is negligible within the particles, which are assumed to be highly porous ($\sim$66\%, with particle density of 1~$\text{g cm}^{-3}$), compared with their drift timescale. The released water vapor diffuses in both directions away from the PDL, but the fraction diffusing outward condenses again and creates an enrichment peak in ice at the location of the snowline. 

\begin{figure*}
\resizebox{\hsize}{!}{\includegraphics[angle=0]{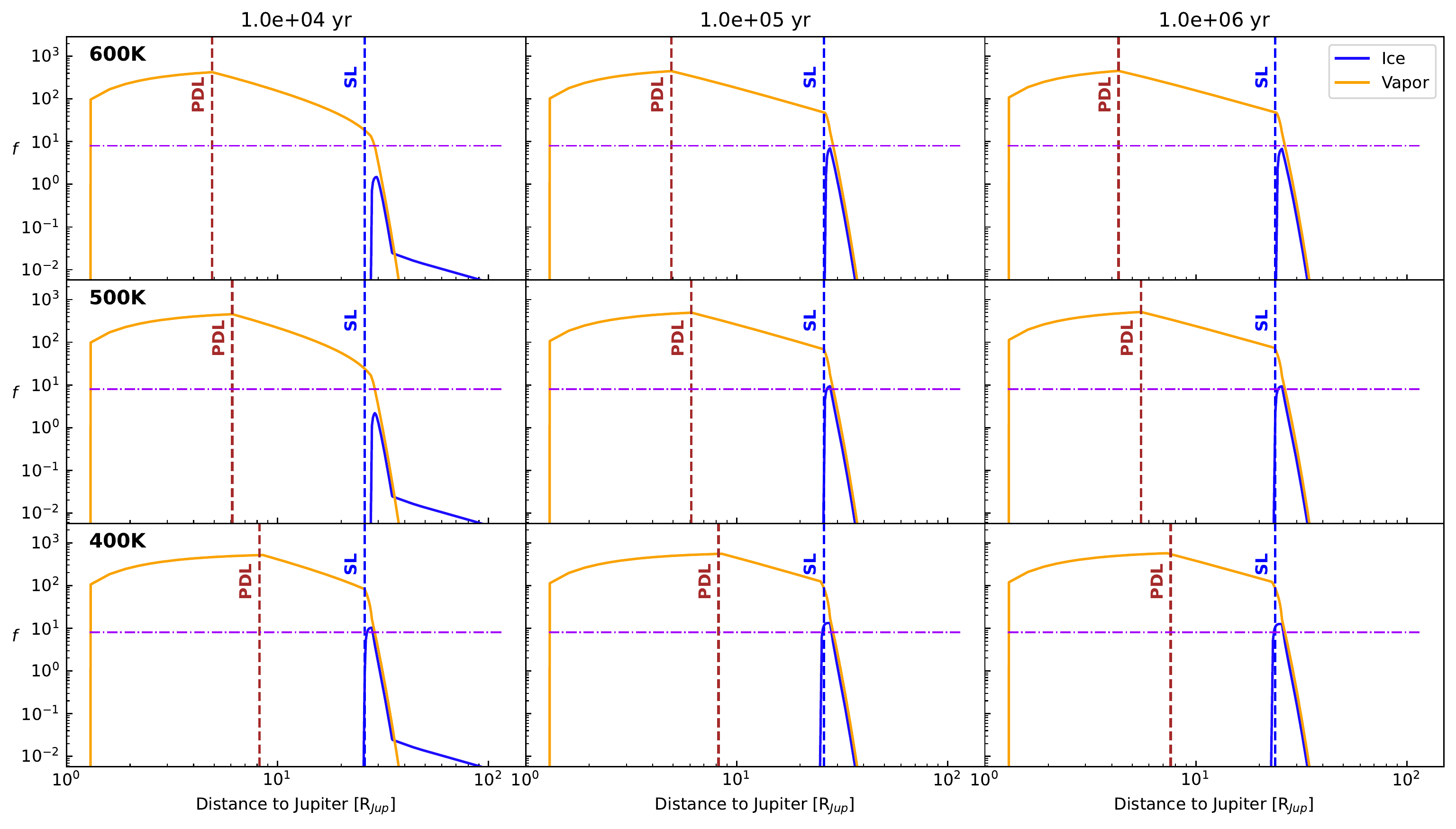}}
\caption{Radial profile of the water abundance within the Jovian CPD expressed as a function of the initial abundance in phyllosilicates, assuming its vapor form is sourced by the dehydration of phyllosilicate-rich particles during their inward drift, and calculated at $t$ = 10$^4$, 10$^5$, and 10$^6$ yr. Three dehydration temperatures are considered: 600 K (top), 500 K (middle), and 400 K (bottom). The horizontal dashed line denotes the threshold at which the amount of condensed water is sufficient to account for that in Ganymede and Callisto. The brown and blue vertical dashed lines correspond to the locations of the PDL and SL in the CPD, respectively.}
\label{fig2}
\end{figure*}

Thermodynamic calculations performed at equilibrium conditions relevant to our CPD model indicate that phyllosilicates could dehydrate at temperatures as low as $\sim$210 K (see Appendix). On the other hand, these calculations do not consider the kinetics of dehydration, for which available experimental data of direct relevance are scarce. Some experiments on the dehydration of phyllosilicates have been carried out under vacuum on serpentine from the Murchison CM chondrite, showing that decomposition starts at temperatures as low as 570 K \citep{Ak92} after heating during 168 hours. The same study indicates the presence of faint dehydration patterns possibly observed at 523 K after heating during 330 h. Because the experimental evidence is limited, and because the timescale of these experiments is only counted in hours, we considered dehydration temperatures of 400, 500, and 600 K within the Jovian CPD, which remain well above the calculated equilibrium temperature.

Figure 2 shows that water vapor diffuses both inward and outward from its source initially located at $\sim$8.2, 6.1, and 4.9 $R_{Jup}$ for the respective dehydration temperatures of 400, 500, and 600 K. Water vapor evolves outward until it freezes at the snowline location. At $t$~=~10$^4$~yr, the water vapor abundance profile already rises by several orders of magnitude compared with its initial content in phyllosilicate-rich particles. An abundance peak of solid ice is already forming at the snowline location, but with a limited extent, assuming dehydration temperatures of 500 and 600 K. At $t$ = 10$^5$ yr, the water vapor distribution remains almost identical, and its outward diffusion continues to induce the formation of water ice in the vicinity of the water snowline. At this epoch, the ice abundance profiles peak at values that are $\sim$13.3, 9.3, and 7.0 times higher than the initial abundance in dust, assuming dehydration temperatures of 400, 500, and 600 K, respectively. After 10$^6$ yr of CPD evolution, the ice abundance profiles slightly decrease and peak at $\sim$12.6, 9.2, and 6.7 times the initial abundance in dust, when considering dehydration temperatures of 400, 500, and 600 K, respectively. Ice peaks are higher at lower dehydration temperatures because the diffusion of vapor takes place over narrower distance ranges between the PDL and the snowline. Figure 2 also shows that the abundance of solids located beyond $R_c$ quickly decreases with time, illustrating the fact that this CPD region loses its solids as pebbles form from the agglomeration of grains and drift inward.

Enrichments higher than 8 times the 10 wt$\%$ of water assumed in phyllosilicates are easily achieved over the first hundreds of thousands of years of the CPD evolution, for dehydration temperatures of 500 K or below. This minimum enrichment factor, combined with the water already present in phyllosilicates (80 + 10 wt$\%$), is enough to explain the ice--to--rock ratio of $\sim$1 estimated in Ganymede and Callisto \citep{So02}, provided that the particle flux delivered to the CPD was constant over the same time period, and comprising phyllosilicates with 10 wt\% water bound as hydroxide groups in the minerals, i.e., an equivalent 90 wt\% of dry minerals. Higher dehydration temperatures do not allow the formation of enough solid ice, unless the interiors of Ganymede and Callisto are rockier than expected \citep{Ne20}. 

\section{Discussion} 
\label{sec:dis}

Most of the current formation scenarios of the Galilean moons (hereafter water-rich scenarios) advocate growth of embryos during their migration in the CPD until the three inner moons Io, Europa, and Ganymede became trapped in mutual mean-motion resonances \citep{Ca02,Sa10,Pe02,Ob20,Ma21}. The outermost moon Callisto would have ended its migration prior to reaching the Laplace resonance because of the removal of the CPD due to photoevaporation \citep{Ob20}, or due to divergent migration resulting from tidal planet-satellite interactions \citep{Ma21}. Mostly based on CPD models similar to the one used in the present work, these water-rich scenarios advocate the migration and growth of Ganymede and Callisto from solids originating from the region extending beyond the snowline, and possessing ice fractions equivalent to those inferred for their interiors. In those scenarios, Io and Europa would have accreted from the same solids, which devolatilized during their inward drift through the snowline \citep{Ca02,Ro17,Sa10}. The formation conditions imposed upon Ganymede and Callisto by our water-depleted CPD are not very different from the water-rich scenarios. Pebbles and satellitesimals accreted by the two forming moons would mostly originate from the  region extending over several Jovian radii beyond the snowline to reproduce their ice-to-rock ratios, as shown by Fig. 2. In our approach, the two inner moons Io and Europa would have grown from the direct accretion of ice-free material in the region interior to the snowline. Europa's hydrosphere would result from the release of the water fraction contained in accreted phyllosilicates \citep{Me21}. The near absence of volatiles in Io's interior would be the consequence of its growth from dehydrated minerals in the region interior to the PDL, or, if it  accreted from the same hydrated minerals that formed Europa, {the loss of its hydrosphere} via hydrodynamic escape \citep{Bi20}.

 The formation of the Galilean moons in a water-depleted CPD is also consistent with the possibility of finding a supersolar abundance of water in Jupiter. The heavy elements accreted in Jupiter's envelope could have been accreted in the form of solids and/or vapors beyond the PSN snowline \citep{Mo21,Ag22} while the CPD would have started to be active when the planet migrated to a water--depleted region inward of the snowline \citep{Al14}. In this context, it is also not excluded that the heavy elements present in the envelope would come from the erosion of an ice-rich core \citep{Al05,Mo17}.
 

Remote sensing or in situ instruments aboard the James Webb Space Telescope and the future JUICE and Europa-Clipper spacecrafts could have, in principle, the capability of discerning the formation scenario elaborated here from those proposed so far in the literature. For example, traditional formation scenarios proposing that the Galilean moons accreted directly from ice-rich solids originating from the PSN predict that Europa, Ganymede, and Callisto should have constant and supersolar deuterium-to-hydrogen (D/H) ratios in water, and typically close to those measured in comets \citep{Ho08,Wa09}. In contrast, a scenario proposing that Ganymede and Callisto accreted from building blocks condensed in an initially warm and dense CPD would display D/H ratios close to the protosolar value \citep{Ho08}. Also, in case of atmospheric loss driven by the accretion heating of Europa, the resulting D/H ratio in its icy crust should be up to 10--100 times higher than the values measured in Ganymede and Callisto, with smaller enrichments for rapid blowoff than for more diffusive escape \citep{Bi20}. Our model predicts that the D/H ratio in Europa would be similar to the value measured in carbonaceous chondrites, which is close to Earth's ocean water \citep{Al12} and that measured in some comets \citep{Bo17}. Our model implies that the D/H ratio in Ganymede and Callisto should be smaller or at most equal to Europa's value because water vapor could isotopically exchange with the CPD's H$_2$ during its outward diffusion toward the snowline. How much lower the D/H ratio is in Ganymede and Callisto, compared to Europa would then strongly depend on the initial thermodynamic structure of the Jovian CPD. Our model finally predicts that the icy phase embedded in Ganymede and Callisto should be by far dominated by water, the icelines of the other volatiles being located at greater distance from Jupiter in the CPD. Such a prediction would be at odds with an Enceladus-like composition, which is close to that observed in comets \citep{Bo17}. Another implication is that the isotopic composition of any detectable nitrogen in the Galilean moons would be more similar to Earth's (largely inherited from organic matter) than Titan's (largely inherited from ammonia).
 
Finally, to be assessed, our scenario highlights the need of additional experiments investigating the kinetics of phyllosilicate dehydration at pressure-temperature condittions relevant to those of the CPD. 
Also, the scenario presented here, as well as the one proposing that Ganymede and Callisto grew from ice-rich solids coming from the PSN, are two endmember scenarios. We then cannot formally exclude that the two moons formed from the combination of both mechanisms.\\

The authors gratefully thank to the Referee for the constructive comments and recommendations. The authors are grateful to Vassilissa Vinogradoff for useful discussions. O.M. acknowledges support from CNES. The project leading to this publication has received funding from the Excellence Initiative of Aix-Marseille Universit\'e -- A*Midex, a French ``Investissements d'Avenir programme'' AMX-21-IET-018. This research holds as part of the project FACOM (ANR-22-CE49-0005-01\_ACT) and has benefited from a funding provided by l’Agence Nationale de la Recherche (ANR) under the Generic Call for Proposals 2022. 

\newpage

\appendix

A lower limit on the dehydration temperature of phyllosilicates can be obtained by determining the temperature at which the following reaction would be at thermodynamic equilibrium in the Jovian CPD:

\begin{equation}
\begin{split}
\rm Mg_3Si_2O_5(OH)_4 (serpentine; chrysotile/lizardite) \rightarrow Mg_2SiO_4 (forsterite) \rm + 0.5Mg_2Si_2O_6 (enstatite) + 2H_2O(g).
\end{split}
\label{Eq_serp}
\end{equation}
								
\noindent Serpentine is selected because it is the most abundant hydrated mineral in chondrites \citep{Br06}; we know that serpentine was present in the early solar system. If the minerals in Eq. \ref{Eq_serp} are approximately pure, which is appropriate when working on a log scale (see below), then the equilibrium constant is given by

\begin{equation}
K \simeq p^2_{H_2O},
\label{Eq_eq}
\end{equation}

\noindent in which it is assumed that the fugacity and partial pressure of H$_2$O (in bars) are interchangeable at the very low pressures that are expected in the Jovian CPD (Fig. \ref{fig1}).

We used the code SUPCRTBL \citep{Zi16,Jo92} to calculate equilibrium constants for Eq. \ref{Eq_serp} as a function of temperature ($T$) at 1 bar total pressure (which is sufficiently close to zero so as to be applicable to disk pressures in our model). Figure \ref{fig3} shows the computed dehydration curves of chrysotile and lizardite polymorphs of serpentine, plotted as log $p_{\rm H_2O}$ vs. $T$. Lizardite is slightly more stable (5--6 kJ/mol) than chrysotile, which is itself considered as a metastable mineral \citep{Ev04}. The sublimation curve of water ice \citep{Wa11} is also shown in Fig. \ref{fig3}. Because the serpentine dehydration curves lie to the right of the ice sublimation curve, it can be inferred that no ice would be present at the conditions of serpentine dehydration.

\begin{figure}
\resizebox{\hsize}{!}{\includegraphics[angle=0]{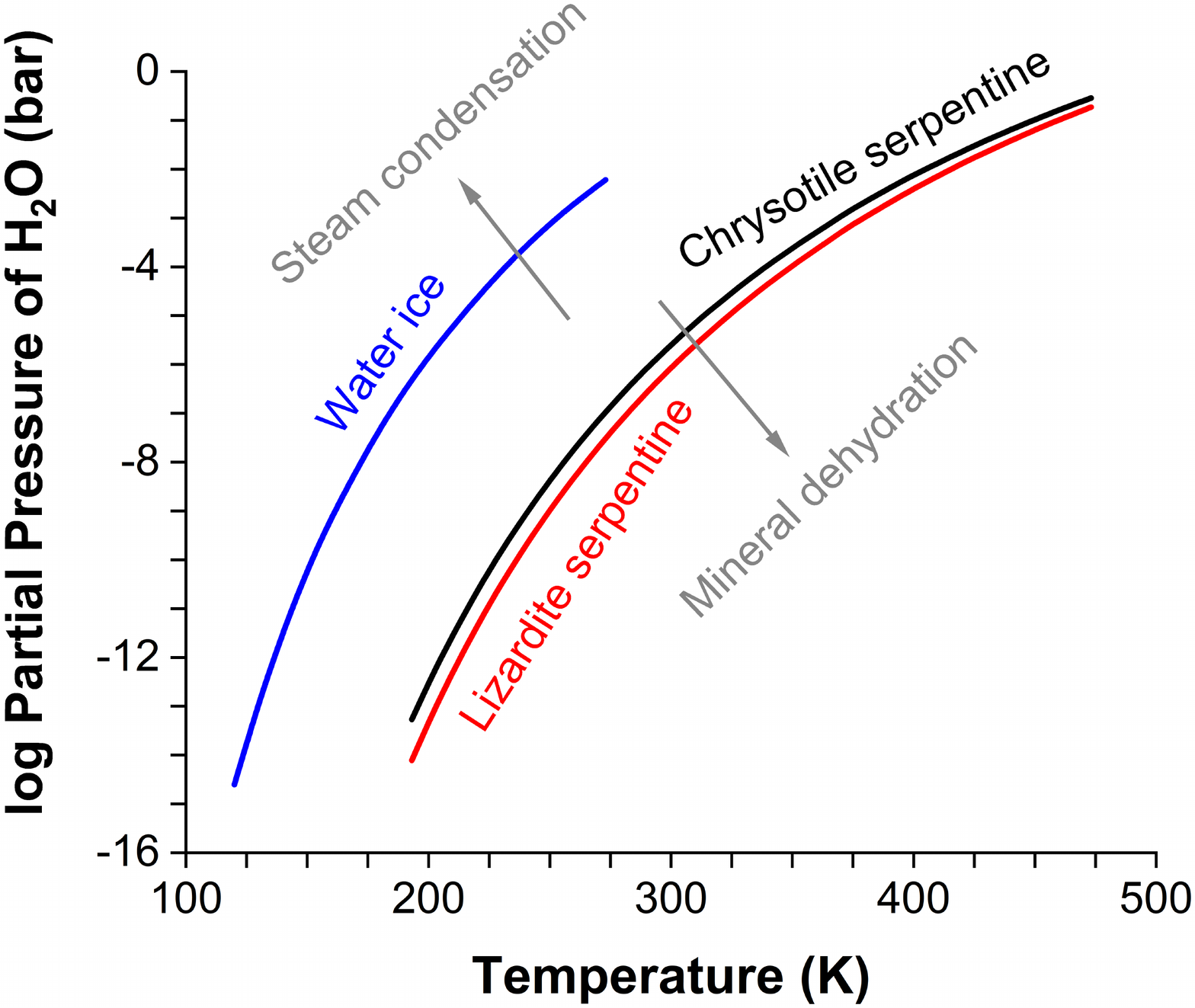}}
\caption{Equilibrium dehydration curves of the serpentine polymorphs chrysotile and lizardite, along with the vapor pressure curve of water ice (Ih). Grey arrows represent examples of how changes in temperature and/or partial pressure of water vapor would make the indicated processes thermodynamically favorable.}
\label{fig3}
\end{figure}

We can recast Eq. \ref{Eq_eq} in terms of the total gas pressure by performing a simple mass balance calculation for the distribution of oxygen in solar composition material \citep{Lo21}. In this canonical calculation, we utilize the approximation that only H, He, C, O, Mg, and Si are significant to the oxygen balance. We assume that Mg and Si exist as their oxide components (MgO, SiO$_2$) in silicate minerals, and the dominant reservoirs of carbon, oxygen, and hydrogen are CO, H$_2$O, and H$_2$, respectively. Because ice cannot be present at the conditions of serpentine dehydration (see above), we assume that all H$_2$O would be present as steam along the serpentine equilibrium dehydration curves. We can then calculate the mole fraction of H$_2$O in the gas phase ($y_{H_2O}$~=~3~$\times$10$^{-4}$). Lastly, we write Eq. \ref{Eq_eq} in the following form:

\begin{equation}
\rm log~\it P_{tot} \simeq \rm 0.5~log~\it K_1~-~\rm log~\it y_{\rm H_2O},
\label{Eq_frac}
\end{equation}

\noindent where $P_{tot}$ refers to the total gas pressure, and $K_1$ is the equilibrium constant at one bar. Figure \ref{fig4} shows the equilibrium dehydration curves of chrysotile and lizardite in $P_{tot}-T$ space. Also shown are condensation curves of water ice for the nominal solar composition model, and for a case in which phyllosilicate dehydration and diffusive transport of water vapor have locally enriched the abundance of water vapor by one order of magnitude. These curves can be compared to the $P_{tot}-T$ profile at 10$^5$ yr from our CPD model (the profiles at 10$^4$ and 10$^6$ yr in Fig. \ref{fig1} are almost identical with respect to the range of $P_{tot}$ in Fig. \ref{fig4}). The thermodynamic limit for the dehydration temperature occurs where the CPD profile crosses the dehydration curve; i.e., at $\sim$206 K for chrysotile and $\sim$215 K for lizardite. These values are insensitive to the assumed speciation of carbon in the CPD. For example, if we adopt an endmember in which no oxygen is bonded to carbon (e.g., C-rich organic matter), then the derived temperatures would increase by only 5--6 K. Strictly speaking, these temperatures are lower limits because it is presently unclear if the rates of dehydration would be fast enough at these temperatures. Our suspicion based on the limited relevant literature (see main text) is that a higher temperature is most likely needed, which is why we assumed in the main text that the temperature should be at least 400 K for phyllosilicate dehydration to serve as a significant source of free water in the Jovian CPD. 

\begin{figure}
\resizebox{\hsize}{!}{\includegraphics[angle=0]{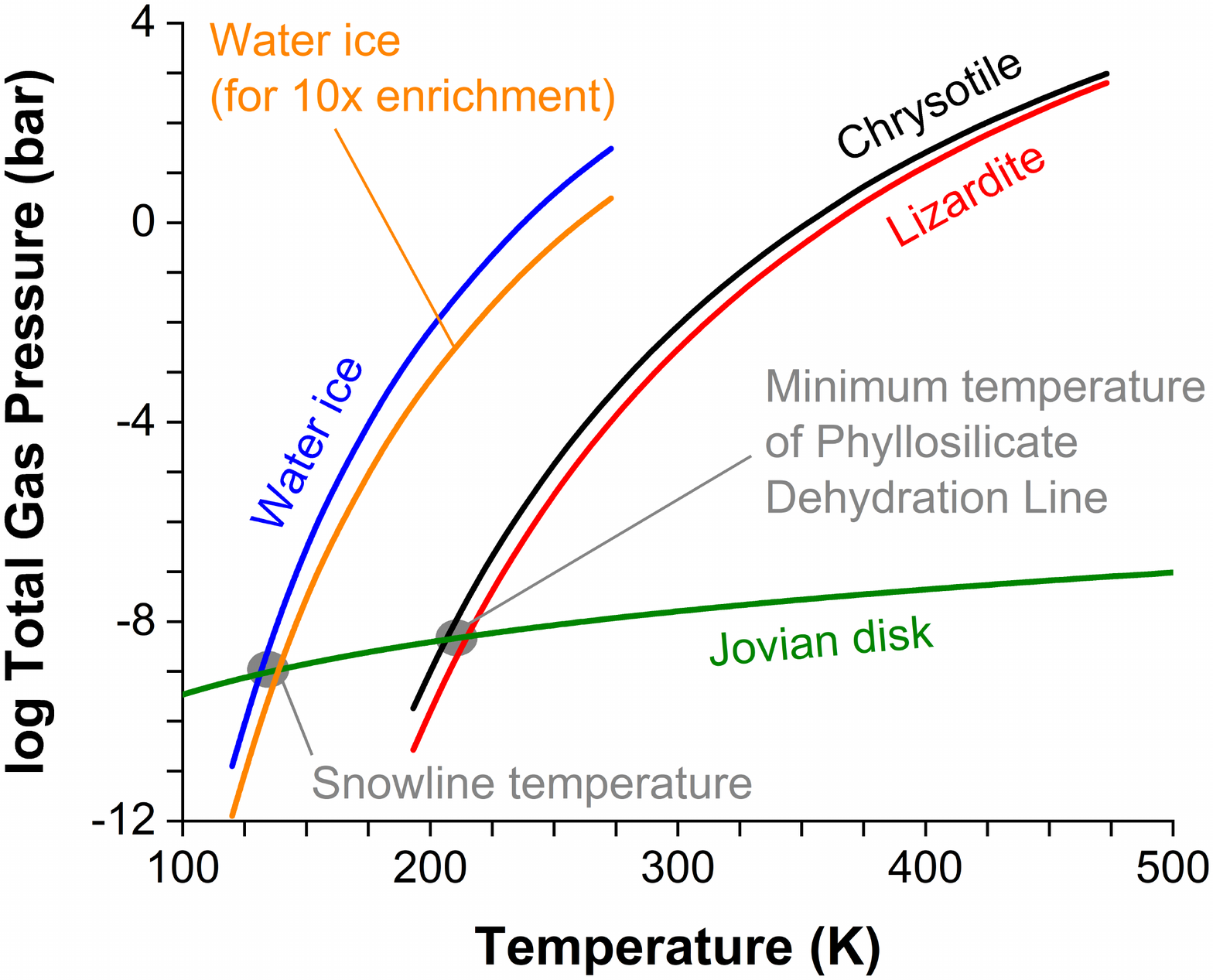}}
\caption{Equilibrium dehydration curves of the serpentine polymorphs chrysotile and lizardite in the presence of solar composition material, where the carbon speciation is dominated by carbon monoxide. Water ice (Ih) condensation curves are shown for (blue) a solar composition case in which $\sim$33\% of the water budget is in phyllosilicates, while the rest is present as water vapor; and for (orange) a case in which the water vapor abundance has been enriched by a factor of 10 relative to the preceding case. The green curve depicts the $P_{tot}-T$ profile from our CPD model after 10$^5$ yr. At temperatures higher than the temperature of curve crossing, dehydration of serpentine is thermodynamically favored. Water vapor will condense as water ice at disk temperatures below the snowline temperature.}
\label{fig4}
\end{figure}


\bibliography{MS}{}
\bibliographystyle{aasjournal}



\end{document}